\documentclass[10pt,pre,twocolumn,aps,superscriptaddress]{revtex4-1}
\usepackage{epsfig,amsmath,amssymb,graphicx,color,calc,epstopdf}

\let\a=\alpha \let\b=\beta \let\g=\gamma \let\d=\delta
\let\e=\varepsilon   
    \let\p=\pi
\let\s=\sigma \let\t=\tau \let\f=\varphi 
   
\let\D=\Delta   
\let\Si=\Sigma   
\let\ee=\epsilon  \let\th=\theta \let\io=\infty

 \def\HH{{\cal H}}

\def\GG{{\cal G}} \def\SS{{\cal S}}
  
\def\ZZ{{\cal Z}}

  \def\erf{\text{erf}}

\def\to{\rightarrow}

\newcommand{\beq}{\begin{equation}}
\newcommand{\eeq}{\end{equation}}
\newcommand{\ba}{\begin{align}}
\newcommand{\ea}{\end{align}}

\newcommand{\oz}{\overline{z}}

\def\SI{Appendix}

\begin{document}

\title{Universal microstructure and mechanical stability of jammed packings}

\author{Patrick Charbonneau}
\affiliation{Department of Chemistry, Duke University, Durham,
North Carolina 27708, USA}
\affiliation{Department of Physics, Duke University, Durham,
North Carolina 27708, USA}

\author{Eric I. Corwin}
\affiliation{Department of Physics, University of Oregon, Eugene, Oregon 97403, USA}

\author{Giorgio Parisi}
\affiliation{Dipartimento di Fisica,
Sapienza Universit\'a di Roma,
INFN, Sezione di Roma I, IPFC -- CNR,
P.le A. Moro 2, I-00185 Roma, Italy
}

\author{Francesco Zamponi}
\affiliation{LPT,
\'Ecole Normale Sup\'erieure, UMR 8549 CNRS, 24 Rue Lhomond, 75005 France}

\begin{abstract}
Jammed packings' mechanical properties depend sensitively on their detailed local structure.
Here we provide a complete characterization of the pair correlation close to contact
and of the force distribution of jammed frictionless spheres.
In particular we discover a set of new scaling relations that connect the behavior of particles bearing small forces and
those bearing no force but that are almost in contact.
By performing systematic investigations for spatial dimensions $d$=3--10, in a wide density range and using
different preparation protocols, we show that these scalings are indeed universal.
We therefore establish clear milestones for the emergence of a complete microscopic theory of jamming. 
This description is also crucial for high-precision force experiments in granular systems.
\end{abstract}


\maketitle

\paragraph*{Introduction --} 
The jamming phenomenon is ubiquitous --  candies~\cite{DCSVSCTC04}, coal~\cite{frenkel:2010}, and colloids~\cite{ZXCYAAHLNY09} all can jam, but its microscopic universality remains debated, even for the most ideal of systems. 
Like any other phase transition, the jamming transition can be approached from the unjammed phase,
e.g. by compressing hard spheres (HS)~\cite{DTS05}, or from the jammed phase, e.g. 
by minimizing the energy of soft spheres (SS)~\cite{OLLN02}.
Yet these two complementary approaches have mostly been developed independently from each other (see~\cite{TS10} for HS and~\cite{LNSW10,vanHecke:2010} for SS).
Unlike standard phase transitions, however, the jamming transition is a non-equilibrium phenomenon that happens deep
inside the glass phase~\cite{MKK08,BJZ11}, and therefore different protocols generate different packings, which may result to conflicting observations.
Indeed, all agree that marginally stable packings of frictionless spheres average $2d$ force-bearing contacts per 
particle~\cite{vanHecke:2010}, but jammed packings' density~\cite{SWM08,KL07,TS10,PZ10,CBS09},  
parts of their microstructure~\cite{SLN06,clusel:2009,TS10}, 
as well as their given name~\cite{TTD00,TS10} are contentious.  
Although 
the jamming ``j'' point was proposed to be unique in the thermodynamic limit~\cite{OLLN02,OSLN03},
there is a growing consensus 
that jamming occurs over a range of
``j'' points~\cite{TTD00,TS10,MKK08,CBS09,PZ10,LNSW10}.
Yet various physical origins have been attributed to the jamming density variation, 
including structural correlations in the initial configuration~\cite{LNSW10},
and the presence of small crystalline regions only detectable by subtle order metrics whose minimization should result
in a single ``maximally random jammed'' state~\cite{TTD00,TS10}.
Others have proposed the intrinsic existence of a range of densities over which packings 
with an identically disordered structure could be found~\cite{MKK08,CBS09,PZ10}. 
A power-law growth of the number of almost-touching particles near jamming has also been identified 
numerically, but different exponents have been found for HS~\cite{DTS05} and SS~\cite{SLN06}. 
If there is microscopic universality, it has yet to fully emerge.

In this letter we bring a different point of view to the problem by systematically investigating how the jamming limit is approached from both sides of the transition
and by varying the dimensionality of space from $d$=3 to 10. This approach allows us to obtain a series of important results.
{\it (1)} Increasing $d\geq 4$ suppresses crystallization~\cite{SDST06,vanmeel:2009b} and the ``spurious'' contribution of
``rattlers''. We can thus show that 
random jammed packings of {\it monodisperse} spheres with identical near-contact structural properties can be obtained 
over a {\it range} of densities (thus confirming results in $d$=3, 4~\cite{TTD00,DTS05,SDST06,CBS09,JST11}), and that this range {\it broadens} with increasing $d$. 
{\it (2)} 
We confirm an earlier suggestion that two exponents $\a$ and $\th$, corresponding to different physical regimes, 
control the mechanical stability of jammed packings~\cite{Wy12}.
The first describes the ``quasi-contact'' regime in which particles are separated by very small gaps $h$,
whose number scales as $h^{-\a}$ for small $h$; 
the second describes the tail of the ``contact'' regime, where the number of particles bearing a small force $f$
scales as $f^\th$.
{\it (3)} We also provide a complete characterization of the microstructure of jammed packings. 
We show that matching the two above regimes provides scaling relations between the exponents and non-trivial scaling functions.
We thus conclude that the mechanical stability of jammed packings is related to their very complex contact microstructure. 
{\it (4)} We find these results to be universal in the sense that they are robust to changes in preparation protocol, packing density, and, in particular, spatial dimension. 

The observation that jammed packings' properties are independent of $d$ suggests that
a mean-field theory should be able to capture the jamming phenomenology~\cite{wyart:2010,goodrich:2012}.
One such treatment, the Gaussian replica theory (G-RT)~\cite{PZ10,BJZ11}, unifies the description of the glass transition 
and of jamming by exploiting an analogy with discrete random optimization problems~\cite{KK07,MKK08}. In this treatment, the HS and SS approaches to jamming
are unified under the assumption that jammed states are the infinite pressure (for HS) or zero temperature (for SS) limit of long-lived metastable
glassy states~\cite{PZ10,BJZ11}.
The theory predicts a growing jamming density range with $d$~\cite{PZ10}, 
the existence of scaling relations for energy and pressure relating the two sides of the jamming transition~\cite{BJZ11},
and makes structural scaling 
predictions that are remarkably satisfied at short distances~\cite{PZ10,BJZ11}. Yet we show here that {\it (5)} G-RT completely fails to describe the structural regime that controls jammed packings' mechanical stability.
Our results {\it (1)-(5)} will thus guide both theory and experiments (through high-precision force measurements~\cite{MSLB07}) towards a better understanding of the jamming transition.

\paragraph*{Packing Generation --}
We consider a system of $N \geq 8000$ identical spherical particles of diameter $\s$ in a fixed volume $V$, under periodic boundary conditions. 
The packing fraction $\f=N V_d(\s/2)/V$, where $V_d(r)$ is the volume of a $d$-dimensional sphere of radius $r$, measures the fraction
of space occupied by particles.
Jammed packings are prepared using two different numerical protocols (see \SI \, for details and reduced units definitions).
\emph{(i)} Approaching jamming from densities below it by Lubachevsky-Stillinger (LS) compressions of HS undergoing Newtonian dynamics while $\sigma$ grows at a fixed rate $\g$=$\dot{\sigma}$~\cite{DTS05}.
The compression, which is tuned to prevent crystallization~\cite{vanmeel:2009b,CIPZ11}, stops when particles are very near contact, defining the packing fraction $\varphi_p^\gamma$ at which the HS reduced pressure becomes infinite. 
\emph{(ii)} 
Approaching jamming from densities above or below it by minimizing the energy $E$ of a 
random configuration of harmonic SS. 
Initial bounds $\s_-$ and $\sigma_+$ that bracket jamming
are evolved iteratively by choosing an intermediate value $\s_{\rm m}$ and
minimizing the energy of the current 
configuration at $\s_+$ (procedure from above) or at $\s_-$ (procedure from below).
The final jammed configurations at the onset of $E\neq0$ have
$\varphi_e^{\downarrow}$ from above and $\varphi_e^{\uparrow}$ from below.
From above, the energy vanishes
with $e=E/N \sim \Delta\f^2$ and the static pressure $P\sim \Delta\f$, where
$\D\f$ is the distance from jamming~\cite{OLLN02}.

\begin{figure}
\includegraphics[width=\columnwidth]{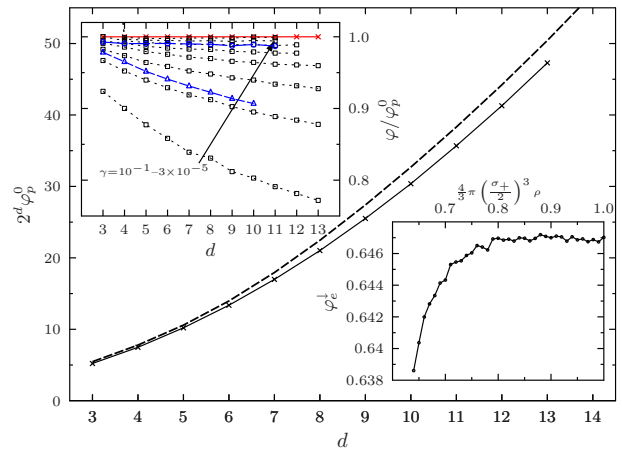}
\caption{
The extrapolated jamming density $\varphi_p^{\gamma\rightarrow 0}$ following the protocol described in Ref.~\cite{CIPZ11} is extended to higher $d$ (solid line and crosses), and
compared with the G-RT prediction for $\f_{\rm GCP}$ (dashed line).
(top inset) The range of jamming densities $\f_p^\gamma$ (squares) is compared to $\f_e^{\mathrm{max}}$ (circles) and $\f_e^{\mathrm{min}}$ (triangles). Note that $\f_e^{\mathrm{max}} \sim \f_p^{\g = 3\times10^{-4}}$ and $\f_e^{\mathrm{min}} \sim \f_p^{\g=3\times10^{-2}}$. (bottom inset) The $d$=3 increase of $\varphi_e^\downarrow$ with $\sigma_+$, in terms of the initial effective packing fraction. 
}\label{fig:phiJ}
\end{figure}

We find the initial $\sigma_\pm$ to have no measurable effect on $\varphi_e^{\uparrow}$.
We formally define $\f_{e}^{\rm min}=\min_{\s_\pm} \f^\uparrow_e(\s_\pm)$, but any reasonable $\s_\pm$ results in the same final density.
By contrast, $\varphi_e^{\downarrow}$  strongly depends on $\s_+$ (Fig.~\ref{fig:phiJ}), but is also independent of $\s_-$. 
We therefore define $\f_{e}^{\rm max} = \max_{\s_\pm} \f^\downarrow_e(\s_\pm)$. 
A practical way of constructing both $\f_{e}^{\rm min}$ and $\f_{e}^{\rm max}$ is to run the energy minimization (respectively from below and from above) 
starting from $\s_-=0$ and $\s_+$ large enough to saturate $\varphi_e^{\downarrow}$ to its maximum. 
Intermediate packing fractions can then be obtained by reducing $\sigma_+$ (Fig.~\ref{fig:phiJ}).
By varying $\s_\pm$ in protocol \emph{(ii)} we can thus construct packings over a density interval 
$[\varphi_e^{\rm min},\varphi_e^{\rm max}]$ that roughly corresponds in protocol \emph{(i)} to $[\varphi_{p}^{\gamma_-},\varphi_p^{\gamma_+}]$ 
with $\gamma_-\approx3\times10^{-2}$ and $\gamma_+ \approx 3\times10^{-4}$ (larger $\gamma$ generate mechanically unstable packings).
The resulting density range is remarkably found to grow steadily from about $2\%$ 
in $d=3$ to nearly $10\%$ in $d=11$ (Fig.~\ref{fig:phiJ}). 
We therefore confirm the similar observation made for $d=3$ binary mixtures~\cite{CBS09}, where the limited available density range and the
subtle crystal order had left some room for debate~\cite{TS10}. 
Note that this range is achieved
by only implementing procedures that compact liquid conﬁgurations.
Ref.~\cite{JST11} has shown that
enlarging the space of procedures enlarges the range of jammed packings, but 
the resulting packings likely have a different microstructure.

The similarity between the jamming density results of the two protocols suggests an underlying physical connection between them. G-RT indeed predicts that packings exist over a finite packing fraction range, whose upper limit is the ``glass close packing'' $\f_{\rm GCP}$~\cite{PZ10}.
By analogy with random combinatorial optimization problems~\cite{KK07}, the densest packing at $\f_{\rm GCP}$ is conjectured to require a
time $\sim \exp(N^a)$ to generate, the exponent being possibly $a \approx (d-1)/d$, based on a nucleation analysis. 
The maximal density that can be reached by the protocols above, which both run in polynomial time in $N$, should therefore be strictly smaller
than $\f_{\rm GCP}$. Figure~\ref{fig:phiJ} shows it to be the case for all $d$, in agreement with G-RT.

\begin{figure}
\includegraphics[width=.8\columnwidth]{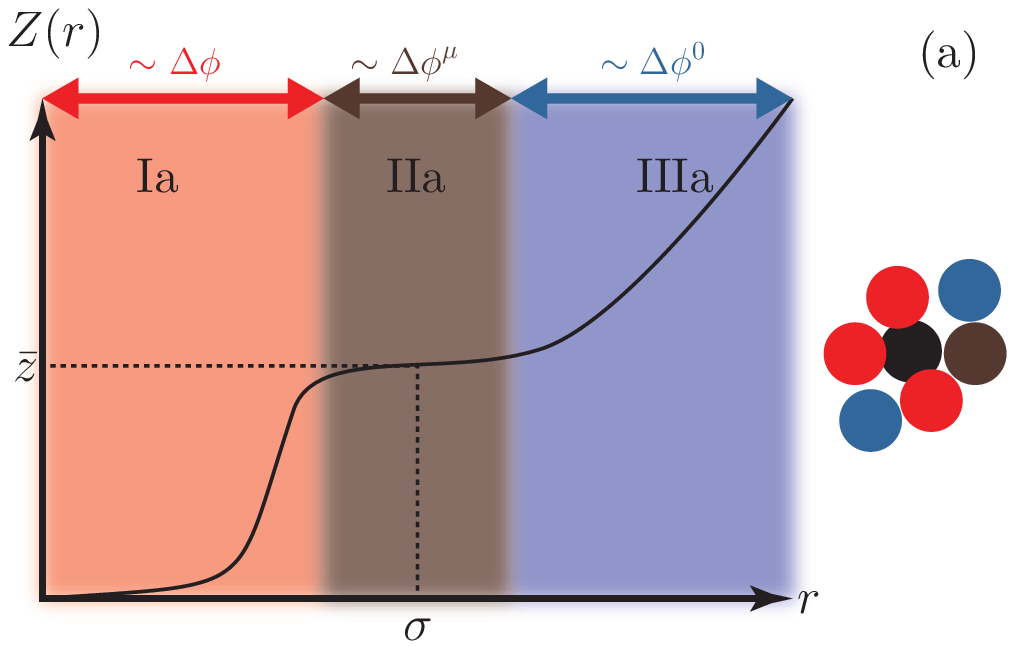}
\includegraphics[width=.8\columnwidth]{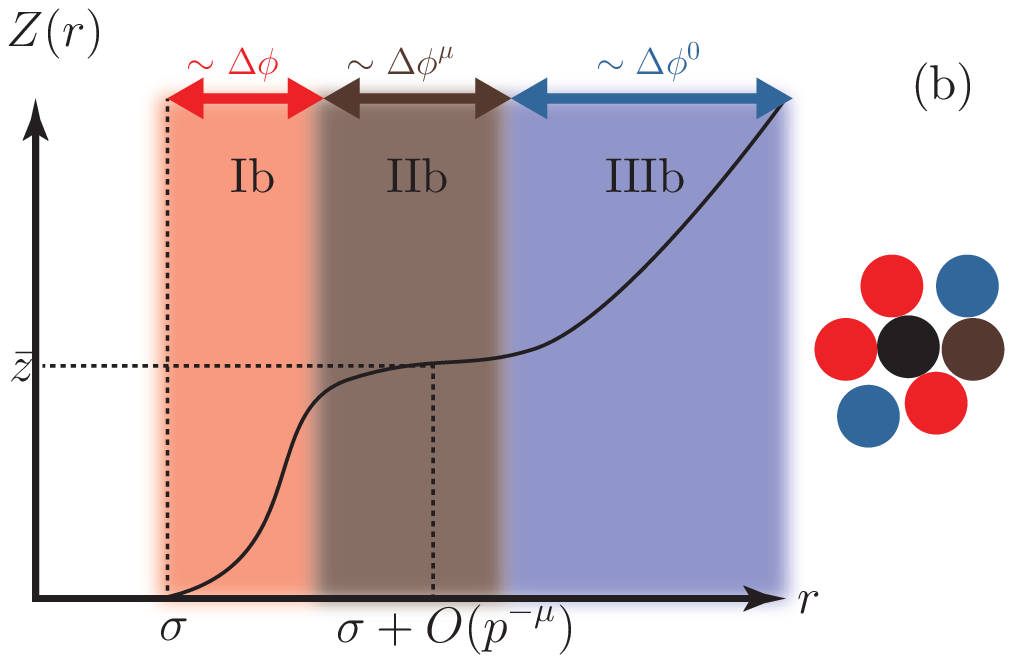}
\caption{
Schematic of $Z(r)$ when approaching jamming from above {\bf (a)} with protocol \emph{(ii)}, and from below {\bf (b)} with protocol \emph{(i)}.
Three distinct scaling regimes can be identified.
The first regime is related to the growth of $Z(r)$ from 0 to $\oz$. It corresponds
to interparticle gaps $h = |r-\s| \sim \D\f$, and hence to
particles that are in contact when $\D\f\to 0$. The last regime corresponds to gaps $h$ that remain finite for $\D \f\to 0$. These particles remain separated at jamming, but these small gaps, $Z(r) -\oz \propto h^{1-\a}$, form a ``quasi-contact'' regime.
The intermediate regime corresponds to gaps $h  \sim \D\f^\mu$. It matches
the two other regimes and disappears when $\D\f\to 0$. 
}
\label{fig:scheme}
\end{figure}

\paragraph*{Scaling functions --}
To determine the universal structure of 
disordered jammed structures, we consider the pair correlation function $g(r)=(\rho N)^{-1}\langle\sum_{i\neq j}\delta(\mathbf{r}+\mathbf{r}_i-\mathbf{r}_j)\rangle$, 
which is the only relevant structural correlation in high $d$ fluids~\cite{FP99}. For numerical convenience, we compute the cumulative structure function
\beq\label{Zr}
Z(r) = \rho \, S_{d-1}  \int_0^r ds s^{d-1} g(s) \ ,
\eeq
where $S_{d-1}$ is the surface of a $d$ dimensional sphere of unit radius.
The function $Z(r)$ thus provides the average number of neighbors within $r$ of a given particle. 
Rattlers are first excluded from the analysis (\SI), but we come back to this point below. 

\begin{figure*}
\includegraphics[width=0.49\textwidth]{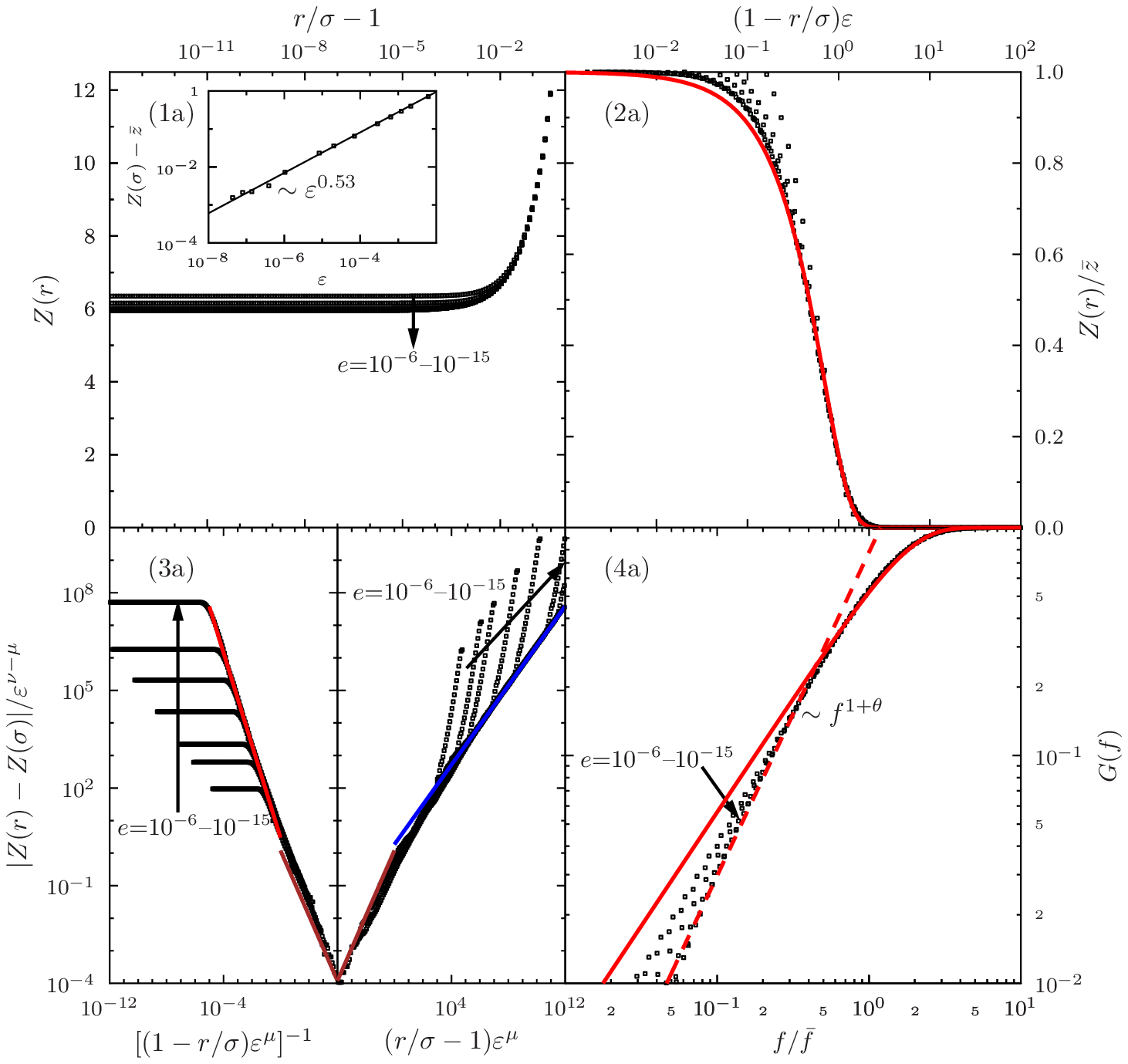}
\includegraphics[width=0.49\textwidth]{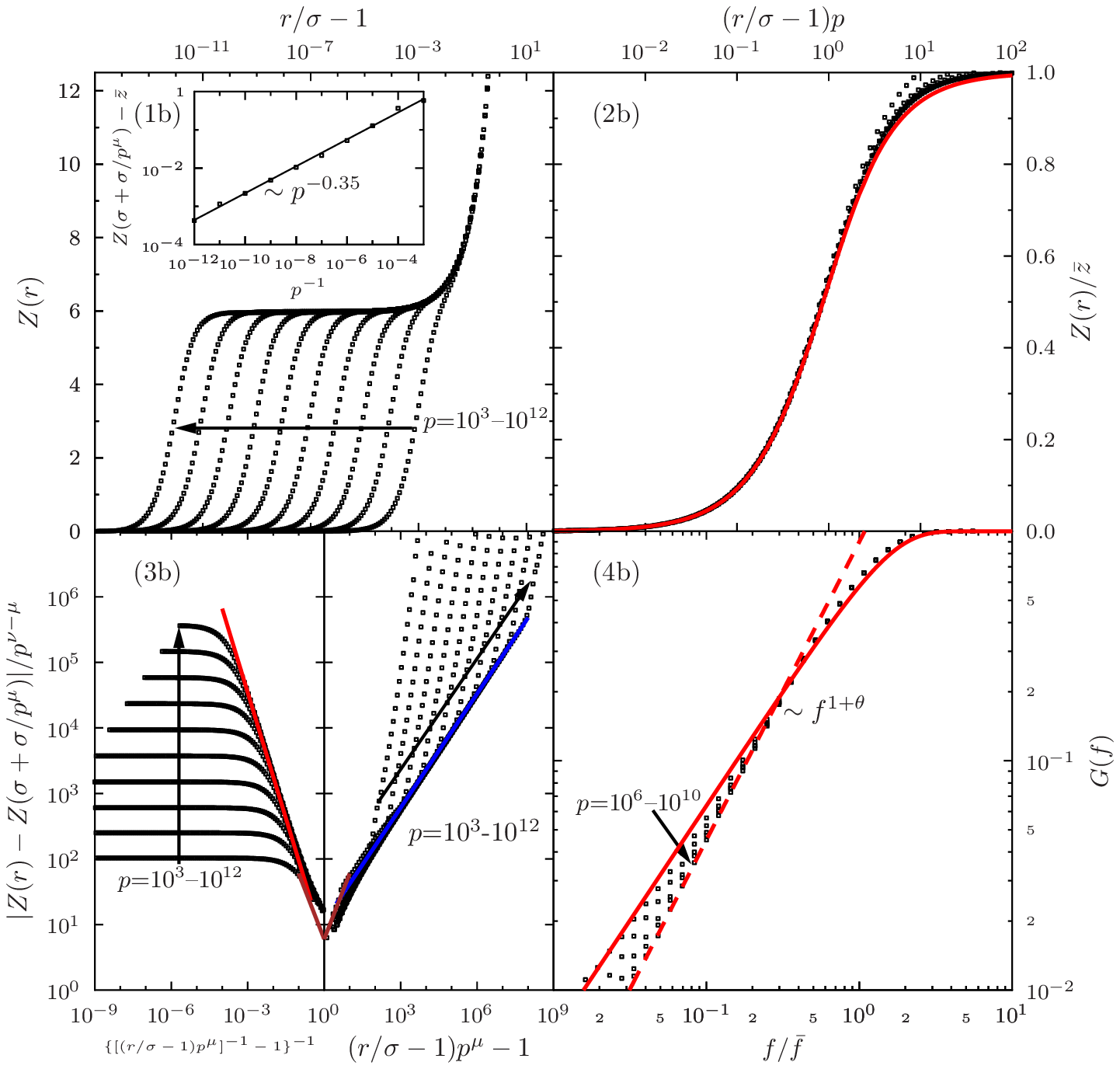}
\caption{
Scaling of the cumulative structure function $Z(r)$ and the cumulative force distribution $G(f)$
in $d=3$ upon approaching jamming from above {\bf (a)} by SS energy minimization (where 
$e \propto|\Delta\f|^2\to 0$ and $\varepsilon=\sqrt{ed/2}$), and from below {\bf (b)} by HS compression (where $p \propto | \D\f|^{-1} \to\io$).
{\bf (1a)} For diminishing $e$, the height of the plateau (inset) converges to the isostatic value with $\zeta$=0.53(3).
{\bf (2a)} The small $r<\sigma$ regime shows the ``contact'' scaling function $\ZZ_+(x)$, which agrees well with the G-RT prediction (red line). 
{\bf (3a)} Rescaling $Z(r)$ using $\mu = (1+\th)/(2+\th-\a)$ and $\nu = \a \mu$ highlights the behavior of the scaling function $|\HH_-(x) - \HH_-(1)|\sim1.2x$ (brown line) along with the $\theta=0.42(2)$ (red line) and the 
$\alpha=0.39(1)$ (blue line) power-law regimes. 
{\bf (4a)} $G(f)$, with power-law tail exponent $\theta=0.42(2)$ (dashed line).
{\bf (1b)} For increasing $p$, $Z(r)$ grows on an earlier scale $r-\sigma \sim p^{-1}$ to a plateau at the isostatic value, whose height (inset) decays with $\zeta=0.36(1)$. 
{\bf (2b)} The small $r-\sigma$ regime shows the ``contacts'' scaling function $\ZZ_-(x)$, which agrees well with the G-RT prediction (red line). 
{\bf (3b)} Rescaling $Z(r)$ using $\mu = (1+\th)/(2+\th-\a)$ and $\nu = \a \mu$ highlights the behavior of the scaling function $|\HH_-(x) - \HH_-(1)|\sim6x$ (brown line) along with the $\theta=0.28(3)$ (red line) and $\alpha=0.42(2)$ (blue line) power-law regimes. 
{\bf (4b)} $G(f)$, with power-law tail exponent $\theta=0.28(3)$ (dashed line), compared with the G-RT prediction (solid line).}
\label{fig:gr_scaling}
\end{figure*}

For both protocols, $Z(r)$ jumps from 0 to a plateau at $\oz$
on a scale 
proportional to the distance to jamming $\D \f$, where
$\oz$ is the isostatic average number of contacts $2d$ plus a correction
$\oz - 2d \propto \D\f^\zeta$ (Fig.~\ref{fig:scheme}). For HS, we find $\zeta$=0.36(2), while $\zeta$=0.53(3) for SS (Fig.~\ref{fig:gr_scaling})~\cite{OLLN02}.
The approach to the isostatic {\it plateau} is characterized by
a long power-law tail 
with exponents $\theta$=0.28(3) for HS and $\theta$=0.42(2) for SS, but the exponent is independent of $d$ for a given model. 
The plateau is extended by a second power-law regime that corresponds to particles in ``quasi-contact'', carrying no force at jamming. 
We find that in this regime the scaling is the same for both
protocols, growing as $Z(r)-\oz \propto (r-\s)^{1-\a}$ with a universal exponent $\alpha=0.42(2)$
until it reaches the trivial large $r$ regime. 
Interestingly, the two power-law regimes can be matched by a scaling function $\HH_\pm$, which defines an additional intermediate regime. 
This intermediate regime shrinks to a point at jamming, but smoothly crosses over from one power-law regime to the other at finite $\Delta\varphi$. 
Consistency therefore sets clear scaling requirements for the different regimes (see \SI\, for scaling analysis)
as detailed in Fig.~\ref{fig:scheme}, and verified in Fig.~\ref{fig:gr_scaling}.

\paragraph*{Force distribution and mechanical stability --}
The consequences of these universal scaling relations on mechanical properties can be gleaned from the probability distribution of inter-particle forces $f$. Here again, we consider
the cumulative distribution $G(f)=\int_0^fP(f')df'$ rather than the pair force distribution $P(f)$, for numerical convenience.

For HS approaching jamming, the average force $\overline{f}\propto p$. In the contact regime the force and distance distributions are also related through a Laplace transform 
(\SI)~\cite{DTS05}. The low-force distribution is thus consistent with $G(f) \propto f^{1+\th}$ and $\th =0.28(3)$. 
For SS approaching jamming from above, the pair potential sets the relation between the force and the pair 
distributions~\cite{OLLN01} (\SI). 
Here again, the low-force tail is consistent with $\th= 0.42(2)$. For both protocols, however, the regime intermediate between 
contacts and quasi-contacts results in deviations from this power-law decay at very weak forces away from jamming.

The large force regime has been thoroughly studied~\cite{OLLN01,OLLN02,OSLN03,corwin:2005,DTS05,vaneerd:2007,radjai:1999}, but the weak force distribution is much less well characterized. Yet it has been proposed by Wyart~\cite{Wy12} that $\alpha \geq 1/(2+\theta)$ is required for mechanical stability.
Both the SS values ($\a= 0.39(1)$, $\th=0.42(2)$) and the HS ones ($\a=0.42(2)$, $\th= 0.28(3)$), however, indicate a slight violation of this condition.
A generalized stability condition of the form $\a \geq (1-\d/2)/(2+\th-\d/2)$~\cite{Wy12} is consistent with our findings for $\d\gtrsim 0.2$, but a direct test of this extended relation is beyond the scope of the current analysis.

\paragraph*{Rattlers --}
Rattlers, i.e., particles with no mechanical contacts, must be considered before concluding that the dimensional and protocol robustness of these results strongly support a universal microscopic description of jamming. Because
their fraction rapidly decreases with increasing $d$ (\SI)~\cite{SDST06}, and their structural contribution is clearly distinct from that of the other particles when $\Delta\varphi\to0$, it is reasonable to remove them from the analysis. 
Rattlers indeed play essentially no role in the scaling regimes in high $d$, while in low $d$, their inclusion introduces noise in $Z(r)$ and $G(f)$ that obscures the scaling relations, which may explain why $\a\approx0.5$ was obtained in Ref.~\cite{SLN06}. Removing the rattlers reveals the robust relationship between microstructure and mechanical properties, in support of jamming having a critical dimension $d_c=2$~\cite{goodrich:2012}. 

\paragraph*{Comparison with microscopic theory --} G-RT, the only available first-principle theory of jammed packings, provides predictions for the contact regime scaling function
$\ZZ_{\pm}(x)$~\cite{PZ10,BJZ11} (\SI). 
We find the form of $\ZZ_{\pm}(x)$ to be extremely accurate when $x$ is of order 1, 
but G-RT fails to capture the ensuing power-law regimes 
(Fig.~\ref{fig:gr_scaling}). G-RT indeed predicts an exponent $\th=0$ for both protocols, and completely misses
the power-law divergence related to $\a$, predicting $\a=0$. A similar deviation is observed at weak forces.
We attribute these discrepancies to the 
Gaussian assumption for the cage form of G-RT, which has recently been found to be erroneous in dense disordered fluids~\cite{lechenault:2010,CIPZ12}.
This non-Gaussian structure also naturally suggests a microscopic explanation for the breakdown of the normal-mode decomposition of jammed states~\cite{schreck:2011,xu:2010}.
Including a non-Gaussian cage to RT ought to provide a better mean-field understanding of the jamming phenomenology.

\paragraph*{Conclusions --}
Our results show that the jamming terminology controversy should be resolved by replacing the j-point~\cite{OLLN02} with the j-line~\cite{MKK08,PZ10}, 
and by distinguishing a \emph{range} of maximally random jammed packings from their partially crystallized counterparts~\cite{TTD00,TS10,JST11}. They also reveal 
that the contacts' complex microstructure in jammed packings is characterized by universal, well-defined scaling
regimes and by their corresponding scaling functions. We give precise numerical predictions for the scaling exponents, and show that the scaling functions
are related to the force probability distribution.
These specific predictions can be tested in soft matter and granular experiments. A preliminary investigation indeed examined the scaling of the peak of the pair correlation function~\cite{ZXCYAAHLNY09}, but our comprehensive predictions can help experimentalists access the full scaling of $Z(r)$ and $G(f)$. This feat should be possible once a force resolution of $\sim5\%\bar{f}$ is experimentally achieved~\cite{MSLB07}. 

Finally, it is worth noting that the present study was limited to zero temperature $T$ in the sense that no thermal motion is allowed in SS and that for HS the energy interaction scale is infinite compared to $T$. At finite $T$, the jamming transition is blurred~\cite{BJZ11}, but vestiges of the scaling relations should remain visible~\cite{ZXCYAAHLNY09}. 
Future work will detail how temperature and its associated anharmonicities affect the $T=0$ scaling relations identified here~\cite{BJZ11,schreck:2011}.

\paragraph*{Acknowledgments --}
We acknowledge discussions with S.~Torquato and S.~Nagel. The European Research Council has provided financial support through ERC
grant agreement no.~247328. P.C. acknowledges National Science Foundation support no. DMR-1055586.

%

\clearpage

\appendix

\renewcommand{\figurename}{Supplementary Figure}

\section{Numerical simulations}

Molecular dynamics simulations of HS and SS energy minimizations of $N$=$2^{18}$ spheres in $d$=3, $N$=8000 in  $4\leq d \leq 9$, and $N$=$2 \times 2^{14}$ in $d$=10 are performed under periodic boundary conditions.  
For $d=3$, the choice of a very large system is motivated by the need of reducing the statistical noise in the intermediate scaling regime for Fig.~\ref{fig:gr_scaling} of the main text.
The $d>3$ system sizes chosen ensure that even when the system is at its densest the box edge remains larger than $2\sigma$, 
which prevents a particle from ever having two direct contacts with another one. 
There are strong reasons to believe that although relatively small these $N$ nonetheless provide a
reliable approximation of bulk behavior. First, with increasing $d$ the largest diagonals of the simulation box 
are $\sqrt{d}$ larger than the box edge.
Second, correlations of the fluid structure are expected to decrease very quickly with increasing
$d$~\cite{FP99,SDST06}, and correspondingly finite-size effects are reduced.
The validity of these rationalizations, which are consistent with the decorrelation property of high $d$ sphere
packings~\cite{TS10},
have been satisfactorily tested in $d=8$ in Ref.~\cite{CIPZ11}.

\subsection{Hard Sphere LS Compressions}

Event-driven HS simulations are performed at thermostated inverse temperature $\beta$ for spheres of mass $m$, starting from random configurations in the limit $\sigma\to 0$. Time has units of $\sqrt{\beta m\sigma^2}$~\cite{SDST06,CIPZ11}, but all dimensional quantities are expressed in units such that $\b=1$ and $m=1$. 
The reduced pressure $p=\b P/\rho$, with $\rho=N/V$ for pressure $P$, diverges at the jamming packing fraction $\f_{p}^\gamma$ as $p \sim |\Delta\f|^{-1}$ with $\Delta\f = \f - \f_{ p}^\gamma$~\cite{DTS05}. Crystallization in $d > 3$ is strongly suppressed, so access to deeply supersaturated starting configurations can be attained via the slow growth rate $\gamma = \dot{\sigma}=3 \times 10^{−4}$~\cite{SDST06,vanmeel:2009b}. In $d=3$, where the crystallization of monodisperse hard spheres is relatively rapid for moderately high packing fractions, $\gamma=10^{-2}$ is employed up to $p=10^3$, but the slow compression rate is used afterwards. The force between particles in high $p$ configurations is measured from the rate of momentum exchange between pairs of particles in simulations with $\gamma=0$. These measurements are made over at least $10^4$ collisions per particle.

\subsection{Soft Sphere Energy Minimizations}

The harmonic SS energy is $E(X,\s) = \sum_{i>j} v(|\vec{r}_i-\vec{r}_j|)$ with $X=\{\vec r_i\}$ and 
$v(r) =\ee (\s-r)^2 \th(\s-r)$.
Units are chosen such that $\ee=1$. 
We start from random sphere configurations $X_+ = X_-$ and use $\s_-$ and $\sigma_+$ that bracket the jamming point, i.e.,
$E(X_-,\s_-)=0$, $E(X_+,\s_+)>0$. Jamming is identified as the onset of non-zero energy, iteratively determined using a bisection method.
At each iteration, an intermediate value $\s_{\rm m}$ is chosen, and
the energy at $\s_{\rm m}$ is minimized via conjugate-gradient (CG) minimization starting from either $X_+$ (from above) or $X_-$ (from below).
The configuration obtained after minimization  $X_{\rm m}$  then substitutes $(X_+,\s_+) = (X_{\rm m},\s_{\rm m})$ if 
$E(X_{\rm m},\s_{\rm m})>0$, or $(X_-,\s_-) = (X_{\rm m},\s_{\rm m})$ if 
$E(X_{\rm m},\s_{\rm m})=0$. 
The procedure stops 
when either of two conditions is satisfied: 
the energy per particle $E(X_+,\s_+)/N$ falls below $10^{-20}$, corresponding to typical 
overlaps of the order of $10^{-10}$; or the change in $E(X_+,\s_+)$ from one minimization step 
to the next is less than the bound set by double precision arithmetic, i.e., $10^{-8} E$.
The final value of $\s_-$ defines the final packing fraction $\f_{e}^{\rm \uparrow}$ in the procedure from below
and $\s_+$ that of $\f_{e}^{\rm \downarrow}$ in the procedure from above.

\subsection{Rattler Analysis}
\begin{figure}
\begin{center}
\includegraphics[width=0.49\textwidth]{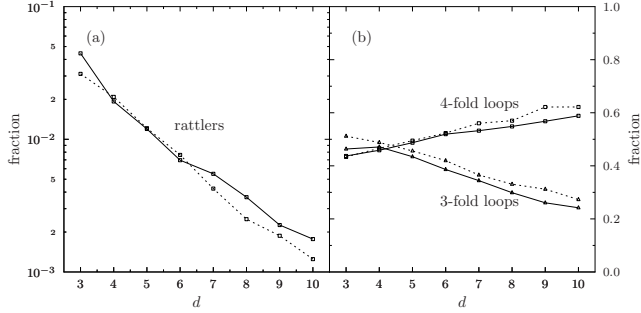}
\caption{{\bf (a)} Fraction of rattlers in HS compressions (dashed line) and in SS energy minimizations from above (solid line). The fraction of particles left outside of the force network, the rattlers, also vanish with increasing $d$. For both protocols, the results suggest their fraction disappears exponentially with $d$. {\bf (b)} Diminishing fraction of 3-member force loops (triangle) and growing fraction of 4-member force loops (square) with $d$ for the two protocols. The force network, which is another observable for comparing the jammed packings, supports their structural similarity. The length of the force loops is also a simple measure of structural correlations. In the mean-field high $d$ limit these loops are expected to become increasingly large, as the structural correlations vanish. The decrease in the fraction of 3-member loops and the growth of the fraction of 4-member loops, accompanied by a constant growth of the average length of the loops, is consistent with this scenario.}\label{fig:loop}
\end{center}
\end{figure}

The rattlers are self-consistently determined by identifying the number of particles with fewer than $d+1$ neighbors within a distance cutoff for the smallest $|\Delta\varphi|$ obtained by a given approach to jamming. In HS, a cutoff of $\sigma (1+100/p)$ is used, which roughly corresponds to a cutoff of $\sigma(1-\varepsilon/100)$ for SS, where the scaling variable $\e = \sqrt{e \, d/2}$. This cutoff slightly overestimates the number of rattlers at jamming, but the rapid diminution of the fraction of rattlers with $d$ (Supplementary Fig.~\ref{fig:loop}) guarantees the robustness of the results.

\subsection{Extraction of the jamming density}
\label{sec:SIprotocols}

\begin{figure}
\begin{center}
\includegraphics[width=0.49\textwidth]{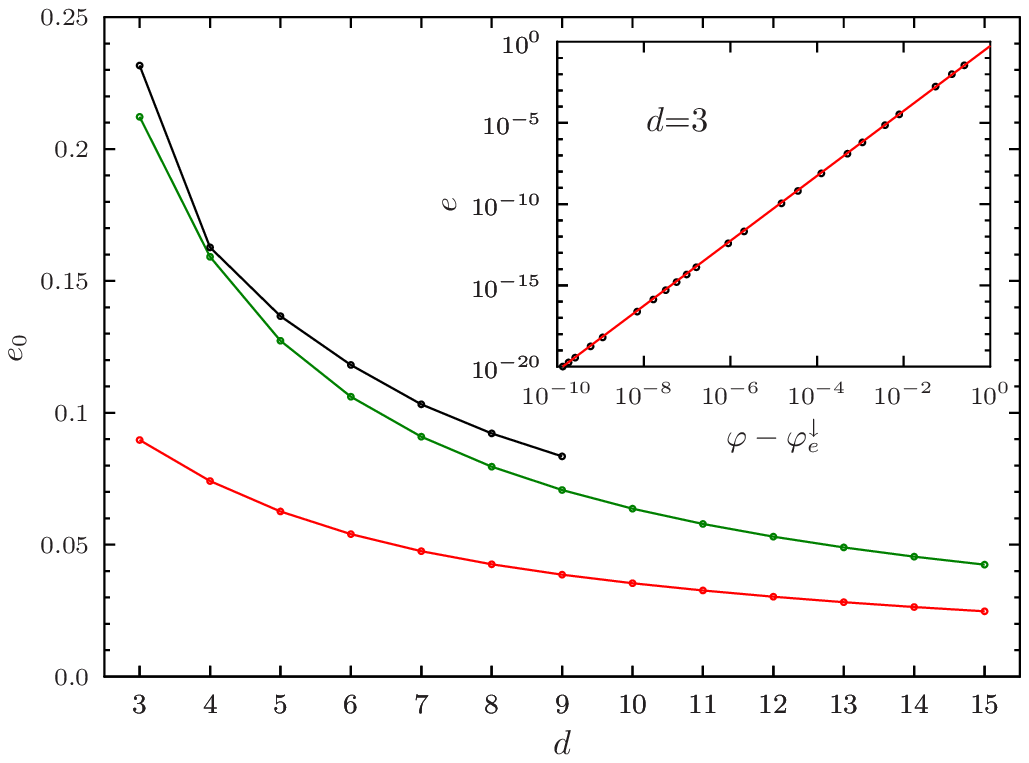}
\caption{
The prefactor of energy $e = e_0 [ (\f - \f_e^\downarrow)/\f_e^\downarrow]^2$ from numerics (black curve) and from G-RT
(red curve, bare data; green curve, corrected data, see Sec.~\ref{app:erep} and~\cite[Section VI.B]{BJZ11}). 
(inset) The scaling $e = e_0 [ (\f - \f_e^\downarrow)/\f_e^\downarrow]^2$ upon approaching $\varphi_e^\downarrow$ from above for $d=3$, with $\s_-=0$ and $\s_+ \to \io$.
Points are numerical data and the line is a quadratic fit used to extract $e_0$ (the value of $\f_e^\downarrow$ is obtained by imposing visually the best alignment of the data).
}
\label{fig:comp}
\end{center}
\end{figure}

Following~\cite{CIPZ11}, for each compression run at fixed $\g$ we perform a linear fit of the line $1/p$ vs $\f$ with $p\gtrsim 50$.
The point where the linear fit vanishes, indicating infinite pressure, defines $\f_p^\g$ for this given run.
Next, for a fixed dimension, we fit $\f_{p}^\gamma = \f_{p}^{\g\to 0} + A \sqrt{\g}$ to
extrapolate the jamming density at $\gamma \to 0$.

For the energy minimization protocol, when approaching jamming from above, the jamming
density can be obtained by accurately fitting the energy data with $e = e_0 [ (\f - \f_e^\downarrow)/\f_e^\downarrow]^2$ (inset in Supplementary Fig.~\ref{fig:comp}).
We focus in particular on minimization  runs performed with initial $\s_-=0$ and $\s_+ \to \io$, which in practice correspond to taking the largest $\s_+$ at which no
variation of $\f_e^\downarrow$ is detected.
The associated prefactor $e_0$ can then be compared with the prediction from G-RT. A good agreement is obtained when the
correction discussed in~\cite[Section VI.B]{BJZ11} is taken into account (Supplementary Fig.~\ref{fig:comp}).

Note that $\f_p^{\gamma\to 0}$ and $\f_e^{\rm max}$ are quite close to each other. 
They indicate the best packing density that can be reached
using our two different compression protocols. Note that according to G-RT
both should be smaller than the maximal packing density of glassy states, called glass close packing $\f_{\rm GCP}$.
According to the theory, it is very unlikely that packings at $\f_{\rm GCP}$ can be produced in polynomial time, hence
it is expected that both $\f_p^{\gamma\to 0}$ and $\f_e^{\rm max}$ are smaller than $\f_{\rm GCP}$.

\section{Scaling functions}
\label{sec:SIscaling}

\subsection{Structure Scaling Analysis}
When approaching jamming with protocol \emph{(i)}, $Z(r)=0$ for $r<\s$, and for $r\geq \sigma$, $p$ parametrizes the scaling function for $Z(r\geq\sigma)$.
A first scaling regime $r-\s \sim p^{-1}$ sees $Z(r)$ grow from 0 to the average number
of ``contacts'' $\oz$ as 
\begin{equation}
Z(r) = \oz \ZZ_-[(r-\s) p /\s ]
\end{equation}
with $\ZZ_-(x) \sim 1 - C x^{-1-\theta}$ when $x\to\io$ for a constant $C$~\cite{DTS05}. 
Force-bearing contacts are only observed at jamming proper, but their signature develops asymptotically.
A second regime for finite $r-\s$ has 
\begin{equation}\label{alphadef}
Z(r) = \oz + C' (r-\s)^{1-\a}\ , 
\end{equation}
where $C'$ is a constant.  At jamming, these nearly touching ``quasi-contacts'' carry no force. 
For large $r$, a trivial regime develops independently of $|\Delta\varphi|$ (Sec.~\ref{sec:SIstructure}). 
Matching the first two scaling regimes implies the existence of an additional intermediate regime $\HH_-$ for
 $r-\s \sim p^{-\mu}$ 
\begin{equation}
Z(r) = \oz + p^{\nu-\mu} \HH_-[(r-\s) p^\mu/\s]
\end{equation}
with $\mu<1$ and $\nu<\mu$. 
Consistency then requires that $\HH_-(x\to 0) \propto -x^{-1-\theta}$ and $\HH_-(x\to\io) \propto x^{1-\a}$ with 
scaling relations $\nu = \a \mu$ and $\mu = (1+\th)/(2+\th-\a)$.

When approaching jamming with protocol \emph{(ii)} from above, the remaining overlaps provide
a scaling variable $\e = \sqrt{e \, d/2}\propto|\Delta\f|$~\cite{BJZ11}. In spite of the very different preparation protocol, similar structural regimes are identified.
In the $r<\s$ contact regime, $Z(r)$ grows from 0 to $\oz$ by a universal scaling function 
\begin{equation}
Z(r) = \oz \ZZ_+[(\s-r) \e^{-1} / \s ]
\end{equation}
with $\ZZ_+(x) \sim 1 - C'' x^{1+\th}$ when $x\to 0$ for a constant $C''$.
For $r>\s$, here again 
\begin{equation}
Z(r) = \oz + C' (r-\s)^{1-\a}\ ,
\end{equation}
hence the two regimes must be matched
by an intermediate scaling function $\HH_+$ for $r - \s \sim \e^{\mu}$
\begin{equation} 
Z(r) = \oz + \e^{\mu-\nu} \HH_+[(r-\s) \e^{-\mu}/\s]
\end{equation}
with  $\mu>1$  and $\nu<1$. 
Consistency here requires that $\HH_+(x \to -\io) \propto  - |x|^{1+\th}$ and $\HH_+(x\to\io) \propto x^{1-\a}$ with $\nu = \a\mu$ and $\mu = (1+\th)/(\a+\th)$. 

In both cases, the exponents $\alpha$ and $\theta$ are determined by collapsing the numerical results using this scaling form, which was repeated for different systems. This constraint leaves a relatively small uncertainty on the final value, which provides the error bar.

\begin{figure}
\begin{center}
\includegraphics[width=0.49\textwidth]{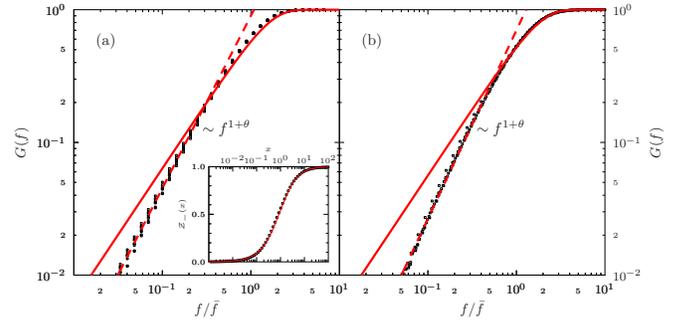}
\caption{
Cumulative force distribution $G(f)$ in $d$=3--10 {\bf(a)} for HS with $\theta=0.28(3)$ and {\bf(b)} for SS with $\theta=0.42(2)$. The force distribution in higher $d$ is essentially the same as in $d=3$ and the high force behavior agrees equivalently well with the G-RT predictions.  The exponents extracted from the small force tail are also numerically indistinguishable. (inset) Test of Eq.~\eqref{eq:FP} from the numerical results for $\ZZ_-(x)$ (points) and plugging the numerical $G(f)$ in Eq.~\eqref{eq:FP} (solid line)
}\label{fig:forcesD}
\end{center}
\end{figure}

\subsection{Force Scaling Analysis}
When HS approach jamming, the cumulative force distribution $G(f)$
approaches a scaling function defined by $G(y \overline{f}) \to \GG_-(y)$.
The relation between the scaling functions $\ZZ_-(x)$ and $\GG_-(y)$ suggested in Ref.~\cite{DTS05}
\beq\label{eq:FP}
\ZZ_-(x) = 1 - x \int_0^\io dy \, \GG_-(y) \, e^{-x y}
\eeq
is verified in Supplementary Fig.~\ref{fig:forcesD}. It follows that if $\ZZ_-(x) \sim 1 - C x^{-1 - \th}$ for $x\to\io$, then $\GG_-(y) \sim y^{1+\th}$ for $y\to 0$, which is here observed for all $d\geq 3$ (Supplementary Fig.~\ref{fig:forcesD}). 
When SS approach jamming from above, the interaction potential gives $f = 2( \s-r )$ for $0\leq r \leq \s$ and zero otherwise, so
$G(f) = 1 - \frac{Z(\s - f/2)}{Z(\s)}$. In the jamming limit $G(2 y \, \e \, \s) \to \GG_+(y) = 1 - \ZZ_+( y )$,
and therefore $\GG_+(y) \sim y^{1+\th}$, as in the previous case. This behavior is here observed in all $d\geq 3$ (Supplementary Fig.~\ref{fig:forcesD}). 

\begin{figure*}
\includegraphics[width=0.49\textwidth]{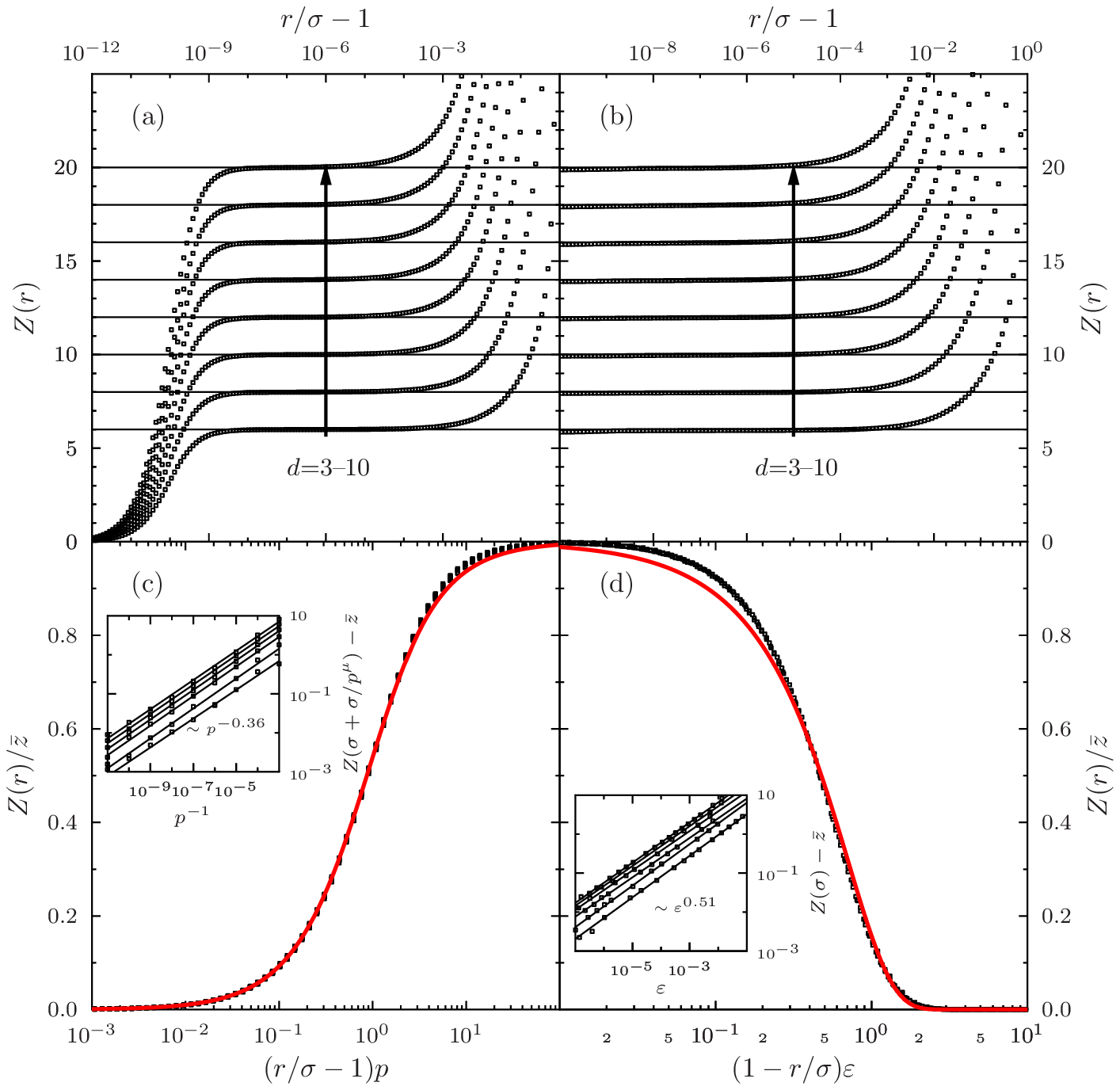}
\includegraphics[width=0.49\textwidth]{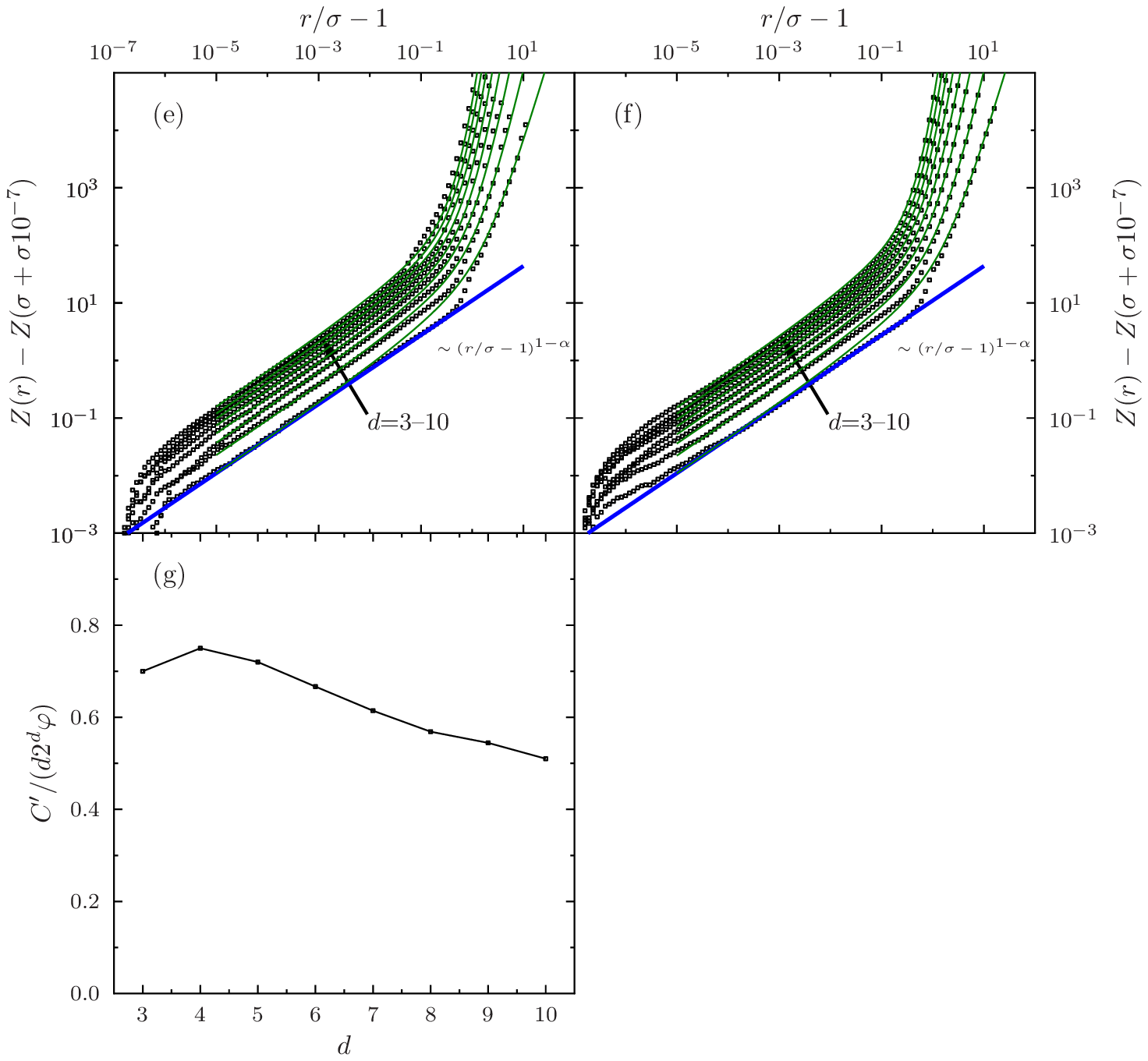}
\caption{Growth of the isostaticity $\bar{z}\sim2d$ (solid lines) plateau with $d$=3--10 for {\bf(a)} HS at $p=10^{10}$ and {\bf(b)} SS at $e\sim10^{-20}$. The HS-SS contact regime {\bf (c)-(d)} collapses remarkably well for all $d$, and the  G-RT predictions (red line) are similarly accurate as in $d=3$. The plateau height also consistently decay from $d$=3 to 8 (insets). {\bf (e)-(f)} The HS-SS quasi-contact power-law growth is also robustly conserved, with a constant $\alpha=0.42(2)$ (blue line). 
The fit to Eq.~\eqref{eq:larger} is also provided (green line).
{\bf (g)} The quasi-contact coefficient $C'$ is such that the region where this regime can be observed shrinks with increasing $d$. 
} \label{fig:Zdelta}
\end{figure*}

\subsection{High $d$ Structure}
\label{sec:SIstructure}

In the contact regime $r -\s \sim \D\f$, 
G-RT predicts that scaling functions $\ZZ_\pm^{\rm RT}$ should describe the growth of $Z(r)$ from 0 to the isostatic value,
as given in Eqs.~\eqref{GRT_HS} and \eqref{GRT_SS}.
Both results are tested in Supplementary Fig.~\ref{fig:Zdelta} for $d$=3--10. The collapse is remarkably good for all $x$. The agreement with the G-RT scaling form is also remarkable for small $x$, but start to deviate from the theoretical prediction when $Z(r)$ approaches the isostatic $\bar{z}\approx2d$ plateau. 
The type of deviation is different from each protocol, but is similar from one dimension to the next for a given protocol. For larger $x$ in the near-contact region, the two protocols robustly produce the same power-law growth, $Z(r)\sim (r-\sigma)^{1-\alpha}$ with $\alpha \sim 0.40(1)$ (Supplementary Fig.~\ref{fig:Zdelta}), which is not predicted by G-RT. 
Because the number of rattlers vanishes with dimension neither of these phenomena can be ascribed to their presence. But because G-RT predictions rely on the individual cages to be Gaussian, which presumably they are not~\cite{CBS09,lechenault:2010,CIPZ12}, it is natural to ascribe the discrepancy to the breakdown of that assumption.

At very large distances, the pair correlation function of any disordered systems trivially has $g(r\gg\sigma)=1$, which corresponds to $Z(r\gg 1)\approx 2^d \varphi [(r/\s)^d - 1]$. 
Unsurprisingly this scaling form captures well the behavior of $Z(r)$ for both protocols and all $d$ at large $r$, but the range of validity also extends with $d$ (Supplementary Fig.~\ref{fig:Zdelta}). 
In order to quantify this effect, we fit the curves of $Z(r)$ for $r>\s$ using the form
\beq\label{eq:larger}
Z(r) = C' (r-\s)^{1-\a} + 2^d \f [ (r/\s)^d - 1) \big] \ .
\eeq
When $d$ grows, the region where the second term is much bigger than the first is
\beq
C' (r-\s)^{1-\a} < 2^d \f [(r/\s)^d - 1) \sim d \, 2^d \f (r-\s)/\s 
\eeq
hence $(r-\s) > [\s \, C'/(d \, 2^d \f)]^{1/\a}$. The fitted values of $C'$ indicate that the crossover point indeed
decreases slowly with $d$.

\section{Replica theory calculations}
\label{sec:SIreplica}

The predictions of G-RT presented in this work are based on earlier results~\cite{PZ10,BJZ11}.
Yet because the calculations in Ref.~\cite{BJZ11} have only been explicitly carried out for $d=3$, and because 
 different observables are considered, additional results are here presented. They are reported in this
section and incidentally provide a somehow simplified derivation of the results of Ref.~\cite{BJZ11}. 
Nonetheless, reading this section requires a detailed knowledge of Refs~\cite{PZ10,BJZ11}, so
the reader who is not interested in the theoretical details can safely skip it.
Note that as in the main text, this section uses reduced units $\ee=1$ and $\s=1$.

\subsection{General expressions}

The approximation scheme used is based on~\cite[Eq.~(22) and (23)]{BJZ11}, which give the replicated free
entropy separated between the harmonic and the liquid contributions 
\beq\label{st1}
\begin{split}
\SS(m,A;T,\f)&= S_h(m,A)+\SS_{\rm liq}(T/m,\f) \\ 
&+ 2^{d-1} \f y_{\rm liq}^{\rm HS}(\f) G(m,A;T) \ , \mathrm{with} \\
G(m,A;T) &= d \int_0^\io dr \, r^{d-1} \, [ q(A,T;r)^m - e^{-\b m v(r)}] 
\end{split}
\eeq
for $m$ replicas at temperature $T$, in a Gaussian cage of variance $2A$.
The function $q(A,T;r) = \int d^dr' \g_{2A}(\vec r')  e^{-\b v(|\vec r - \vec r'|)}$ is defined in~\cite[just after Eq.~(16)]{BJZ11} where $\gamma_{2A}$ is a normalized and centered Gaussian of variance $2A$, and $y_{\mathrm{liq}}^{\mathrm{HS}}$ is the HS cavity function.
Introducing bipolar coordinates, as in~\cite[Appendix C.2.a]{PZ10}, we obtain the generalization of~\cite[Eq.~(C16)]{PZ10}
to the soft sphere potential $v(r) = (1-r)^2 \th(1-r)$
\beq\label{st2}
\begin{split}
q(A,T;r) &= \int_0^\io du \, e^{-\b v(u)} \left(\frac{u}r \right)^{\frac{d-1}2}
\frac{ e^{ -\frac{(r-u)^2}{4A} }}{\sqrt{4\p A}} \times \\
&\times
\left[ e^{ -\frac{ru}{2A} } \sqrt{\pi \frac{ru}{A}} I_{ \frac{d-2}{2} } \left( \frac{ru}{2A} \right)\right] \ .
\end{split}\eeq

The above equations (\ref{st1}) and (\ref{st2}) are the starting point of all the needed replica calculations for our analysis, and
because we focus on the ``jamming limit'' of these equations, we take
$T \to 0$ with $\t = T/m$ and $\a= A/m$ held constant~\cite{BJZ11}.
In Ref.~\cite{BJZ11} this limit was taken using a simplified form of Eq.~(\ref{st2}) for $d=3$,
but we here generalize the calculation to arbitrary $d$.
The crucial observation~\cite[Eq.~(C21)]{PZ10} is that when $z\to \io$, $e^{-z} \sqrt{2\pi z} I_n(z) \to 1$. In the
jamming limit the term in the second line of Eq.~(\ref{st2}) therefore disappears, because $A\to 0$ while $r$ and $u$ are of order 1.
The remaining integral can then be evaluated via the saddle point approximation, because both $\b$ and $1/A$ diverge. 
Consider first the case $r<1$.
Assuming that the saddle point $u^*<1$,
one has to maximize the function
$-\b(1-u)^2 - (r-u)^2/(4A)$ which consistently maximizes 
$u^*=(4\b A+r )/(1+4 \b A) = (4 \a + \t r)/(4 \a + \t)<1$. 
Consider next the case $r>1$. Assuming that in this case $u^*>1$, we have $v(u)=0$ 
and we thus consistently find $u^*=r>1$.
Replacing these expressions for $u^*$ in Eq.~(\ref{st2}) and
taking the jamming limit we obtain
\beq\label{j2}
q(A,T;r)^m \to e^{-\frac{(1-r)^2}{4\a + \t} \th(1-r)} \ ,
\eeq
and Eq.~(\ref{st1}) reduces to
\beq\label{j3}
\begin{split}
\SS_0(\a,\t;\f)&= -\frac{d}2 [ \log(2\pi \a)+1] + \SS_{\rm liq}(\t,\f) \\ 
&+ 2^{d-1} \f y_{\rm liq}^{\rm HS}(\f) G_0(\a,\t) \ , \\
G_0(\a,\t) &= d \int_0^1 dr \, r^{d-1} \, [ e^{-\frac{(1-r)^2}{4\a + \t}}  - e^{- \frac{(1-r)^2}{ \t}}] \ ,
\end{split}
\eeq
which replaces~\cite[Eqs.~(D3) and (D4)]{BJZ11}.

The approach to jamming from above is described by the 
small $\t$ limit~\cite[Appendix D.2]{BJZ11}.
In this limit we can consider the SS Mayer function
as a small perturbation of the HS one and use 
standard liquid perturbation theory to write
\beq\label{j1}
\SS_{\rm liq}(\t,\f) = \SS_{\rm liq}^{\rm HS}(\f) + d \, 2^{d-1} \f y_{\rm liq}^{\rm HS}(\f) \int_0^1 dr \, r^{d-1} e^{-(1-r)^2/\t} \ .
\eeq
Plugging this result in Eq.~(\ref{j3}) we then get
\beq\label{j4}
\begin{split}
\SS_0(\a,\t;\f)&= -\frac{d}2 [ \log(2\pi \a)+1] + \SS^{\rm HS}_{\rm liq}(\f) \\ 
&+ 2^{d-1} \f y_{\rm liq}^{\rm HS}(\f) d \int_0^1 dr \, r^{d-1} \, e^{-\frac{(1-r)^2}{4\a + \t}}  \ .
\end{split}
\eeq
Note {\it en passant} that the cancellation of the second term in Eq.~(\ref{j1}) with a corresponding term in $G_0(\a,\t)$
is not surprising, as stated in~\cite[Appendix D.2]{BJZ11}, but has a deep physical interpretation. Indeed, Eq.~(\ref{j3}) shows
that the ``bare'' SS potential $e^{-(1-r)^2/\t}$ is modified around jamming by the presence of $m-1$ additional  replicas
(with $m\to 0$) that ``renormalize'' it to $e^{-(1-r)^2/(4 \a+\t)}$, as obtained in Eq.~(\ref{j2}). 
The crucial point is that the latter potential does not have
a singularity when $\t\to0$, ensuring a smooth crossover and appropriate scalings around jamming.

\subsection{The energy prefactor}
\label{app:erep}

Starting from Eq.~(\ref{j4}) and repeating the calculations of Ref.~\cite[Appendix D.2]{BJZ11}, we finally obtain the quadratic
scaling of the energy as a function of $\D\f$ when approaching jamming from above. The general expression for the prefactor
is then easily obtained. A further simplification is obtained by assuming that $\a$ is small at the jamming point, and developing
the resulting expressions in powers of $\sqrt{\a}$~\cite{PZ10}. Doing so and optimizing over $\a$ and $\t$, one finally obtains
\beq\label{tutto1}
\begin{split}
\sqrt{\a(\f)} & = \frac{1}{2^d \f y^{\rm HS}_{\rm liq}(\f)} \sqrt{\frac4\pi} \ , \\
\Si_0^{\rm HS}(\f)  &= -d  \log \left( \frac{\sqrt{8}}{2^d \f y^{\rm HS}_{\rm liq}(\f)} \right) + \frac{d}2 + \SS_{\rm liq}^{\rm HS}(\f) \ , \\
\SS_1(\f) &= \frac{d \, \pi}{32} [ 2^d \f y^{\rm HS}_{\rm liq}(\f) ]^2 \ , \\
\t(\f) &= - \Si_0^{\rm HS}(\f) / (2\SS_1(\f)) \ , \\
e(\f) &= [\Si_0^{\rm HS}(\f) ]^2/(4 \SS_1(\f)) = d \, \t^2 / (8 \a) \ .
\end{split}
\eeq
The second line recovers the result of Ref.~\cite[Eq.~(77)]{PZ10}. The glass close packing point $\f_{\rm GCP}$~\cite{PZ10}, which is defined by the complexity $\Si_0^{\rm HS}(\f)=0$,
is reported in Fig.~\ref{fig:phiJ}.
The first line shows that $\sqrt{\a(\f_{\rm GCP})}$ is indeed very small, being $\sim 0.01$ in $d=3$ and decreasing with dimension.
Linearizing the last line around $\f_{\rm GCP}$ and using that $\Si_0^{\rm HS}(\f) $ vanishes linearly, one obtains the quadratic scaling
of the energy and its prefactor~\cite{BJZ11}.
G-RT results in Supplementary Fig.~\ref{fig:comp} have been obtained from this procedure, using the Carnahan-Starling 
equation of state in $d$ dimensions for the HS liquid~\cite[Eq.~(82)]{PZ10}.

A last remark on the energy prefactor is in order. G-RT results in a discrepancy
between the pressure computed from thermodynamics and that computed from the structural (see  Refs.~\cite[Eq.~(89)]{PZ10} and
\cite[Section VII.B]{BJZ11}). This difference might have its
origin in the fact that only two-body effective replica interactions are kept in this treatment. Indeed in the limit $d\to\io$,
where this approximation should be exact, the discrepancy disappears. It has also been observed in Ref.~\cite[Section VI.B]{BJZ11} that
a much better agreement between theory and numerical data is obtained if the distance from jamming $\D\f$ is 
corrected to account for this discrepancy. The correction factor obtained from the theory corresponds to the factor needed
to impose the equality in~\cite[Eq.~(89)]{PZ10}, so rescaling $\D\f$ is equivalent to rescaling $e_0$. The rescaled predictions 
are also reported in Supplementary Fig.~\ref{fig:comp}.

\subsection{Scaling functions}

To complete the analysis we compute the scaling functions $\ZZ_\pm(x)$ predicted by G-RT. Consider first the HS case, approaching jamming from below.
The contact peak of $g(r)$ on approaching jamming is given by~\cite[Eq.~(90)]{PZ10}
\beq\label{grscal}
\begin{split}
g(r) & = g(1) \D_0\left( 2^{d-1} \f g(1)
\frac{\sqrt{\pi}}2 (r-1) \right) \\
& = \frac{p}{2^{d-1}\f} \D_0 \left(  p \frac{\sqrt{\pi}}2 (r-1)  \right) \ ,
\end{split}
\eeq
where $\D_0(x) = 1 - \sqrt{\pi} x e^{x^2} [1 -\erf(x)]$.
The validity of the thermodynamic relation $p = 1 + 2^{d-1} \f g(1)$ 
is here assumed. As we discussed above, this relation is violated by the theory,
but the correction is here unimportant. If one does not want to use this relation, it is
sufficient to replace $p \to  2^{d-1} \f g(1)$, but in the end this substitution does not affect the prediction for the scaling function.

Integrating Eq.~\eqref{grscal} using Eq.~(\ref{Zr}) 
we get, after changing the variable to $y = p \frac{\sqrt{\pi}}2 (s-1)$
\beq\begin{split}
Z(r) & =  2 d \, p \int_1^{r} ds \, s^{d-1} \, \D_0\left( p \frac{\sqrt{\pi}}2 (s-1) \right) \\
& = 2 d \frac{2}{\sqrt\pi} \int_0^{p \frac{\sqrt\pi}2 (r-1)} dy \, \left[ 1 + \frac{y}p \frac{2}{\sqrt\pi} \right]^{d-1} \D_0(y)\ .
\end{split}\eeq
We notice now that $p \sim 1/\D\f$ and $r - 1 \sim \D\f$. The integration is therefore over an interval of order 1. The first term in the integrand can be neglected
because for $y$ of order 1, so this term goes to 1 when $\D\f\to 0$. Finally
we obtain in the contact region
\beq\label{GRT_HS}
\begin{split}
\frac{Z(r)}{2d} &=  \frac2{\sqrt{\pi}} \int_0^{\frac{\sqrt\pi}2 x} dy \D_0(y) \\
&= 1 - e^{\frac\pi2 x^2} \left[ 1- \erf\left( \frac{\sqrt\pi}2 x \right) \right] \equiv \ZZ_-^{\rm RT}(x)\ ,
\end{split}
\eeq
where $x = p  (r-1)$. This prediction is tested in Fig.~\ref{fig:gr_scaling}  of the main text 
and in Supplementary Fig.~\ref{fig:Zdelta}.

Next consider the SS case approaching jamming from above, working in the jamming limit.
$T\to 0$ with $\t = T/m$ and $\a=A/m$.
In this case the calculation starts from~\cite[Eqs.~(17) and (48)]{BJZ11}.
Using bipolar coordinates we can write
\beq\label{Zp1}
\begin{split}
\frac{g(r)}{y_{\rm liq}^{\rm HS}(\f)} &= e^{-\b v(r)} \int_0^\io du \, q(A,T;u)^{m-1} \left(\frac{u}r \right)^{\frac{d-1}2}
 \times \\
&\times\frac{ e^{ -\frac{(r-u)^2}{4A} }}{\sqrt{4\p A}}
\left[ e^{ -\frac{ru}{2A} } \sqrt{\pi \frac{ru}{A}} I_{ \frac{d-2}{2} } \left( \frac{ru}{2A} \right)\right] \ .
\end{split}\eeq
We now need to improve Eq.~(\ref{j2}) by considering the quadratic corrections around the saddle point,
which for $r<1$ leads to
\beq\label{Zp2}
q(A,T;r) \sim e^{-\frac{(1-r)^2}{m(4\a + \t)}}  \left[ \frac{4\a + \t r}{r(4\a+\t)} \right]^{\frac{d-1}2} \frac{1}{\sqrt{1+4\a/\t}}      \ .
\eeq
Plugging Eq.~(\ref{Zp2}) in Eq.~(\ref{Zp1}), dropping as before the last term in square brackets in Eq.~(\ref{Zp1}),
and evaluating the integral via the saddle point approximation including quadratic corrections, we find for $r<1$ that
\beq
\begin{split}
\frac{g(r)}{y_{\rm liq}^{\rm HS}(\f)} &= 
e^{ - \frac{4\a + \t}{\t^2} (r-1)^2 }
\left( 1 +\frac{4\a}\t \right) \times \\ &\times \left[ \frac1{r} \left( 1 + (r-1)\left(1+\frac{4\a}\t\right) \right) \right]^{d-1} \times \\
 & \times \th\left(1 + (r-1) \left(1+\frac{4\a}\t\right) \right)\ .
\end{split}
\eeq
Plugging this result in Eq.~(\ref{Zr}), 
and assuming that we approach jamming from above, so that $\t \ll 4\a$, 
we obtain for $r>\frac{4\a}{\t+4\a} \sim 1 - \frac\t{4\a}$ that
\beq\begin{split}
Z(r) &= y_{\rm liq}^{\rm HS}(\f) d \, 2^d \f \frac{4\a}\t \int_{1-\frac{\t}{4\a}}^r ds \, s^{d-1} \times \\
&\times
\left[ \frac1{s} \left( 1 + (s-1)\frac{4\a}\t \right) \right]^{d-1}e^{ - \frac{4\a}{\t^2} (s-1)^2 } \ .
\end{split}\eeq
Changing variables to $y = (1-s) \frac{4\a}\t$ and using the first line of Eq.~(\ref{tutto1}) gives
\beq
Z(r)= d \, \sqrt{\frac{4}{\pi \a}}  \int_{(1-r) \frac{4\a}\t  }^1 dy \,
( 1 -y )^{d-1}e^{ -\frac{y^2}{4\a} } \ .
\eeq
Although this result is already the desired scaling function, we can further simplify it by noting that $\a$ is small
to write
\beq\begin{split}
Z(r) & =  d \, \sqrt{\frac{4}{\pi \a}}  \int_{(1-r) \frac{4\a}\t  }^\io dy e^{ -\frac{y^2}{4\a} } \\ 
&=2 d \left[ 1 -\erf\left( \frac{2 \sqrt{\a}}\t (1-r) \right) \right] \\
&=2 d \left[ 1 -\erf\left( \sqrt{\frac{d}{2e}} (1-r) \right) \right]  \ , 
\end{split}\eeq
where in the last line we used the last line of Eq.~(\ref{tutto1}).
The resulting prediction
\beq\label{GRT_SS}
\ZZ_+^{\rm RT} = 1 -\erf(x)
\eeq
is tested in Fig.~\ref{fig:gr_scaling} of the main text and in Supplementary Fig.~\ref{fig:Zdelta}.

\end{document}